# A Robust Client Verification in cloud enabled m-Commerce using Gaining Protocol

Chitra Kiran N.
*Research Scholar*:
Dept of Electronics & Communication Engg.
UVCE
Bangalore, India

Dr. G. Narendra Kumar
*Prof*. Dept. of Electronics & Communication Engg.
UVCE
Bangalore, India

*Abstract*— The proposed system highlights a novel approach of exclusive verification process using gain protocol for ensuring security among both the parties (client-service provider) in m-commerce application with cloud enabled service. The proposed system is based on the potential to verify the clients with trusted hand held device depending on the set of frequent events and actions to be carried out. The framework of the proposed work is design after collecting a real time data sets from an android enabled hand set, which when subjected to gain protocol, will result in detection of malicious behavior of illegal clients in the network. The real time experiment is performed with applicable datasets gather, which show the best result for identifying threats from last 2 months data collected.

*Keywords-e-payment, m-Commerce, Cloud Computing, Verification, Security*

## I. INTRODUCTION

commerce is the most recent trend of the technology combined with cloud enabled services where security is one of the most prime issues of any business in recent years. There are three common architecture for establishing the application of clouding computing such as Server as a service (SaaS), Platform as a service (PaaS) and Infrastructure as a service (IaaS) [1] [2] [3] [4]. In order to implement the above application based on cloud computing, the assurance of information security, authentication, and integrity is the most critical foundation [5] [6] [7] [8]. The previous research has already witnessed the use of IPv6 which has tremdeously enhanced the flaws of the IPv4 and repair a lot of security loopholes. But the networking which is of wireless type around the location at any instant of time and location will outsource more data on security issues. So, improving the procedures and techniques for data security in m-commerce is the prime key to control the success of business. With related to the data security of the cloud computing, such problems which are associated with the application and the service of the cloud computing should be deeply concerned e.g. devices security of the user, the existing threats poses towards illegal hacking on mobiles are very crucial [8] which has no solution till now. With the advent of cloud services from prime companies like Google, IBM HP etc, the security is absolutely poorly designed when it comes to mobile commerce applications like micropayments, online transaction etc.

The current mobile commerce is experiencing a very fast growth and also there is also an inclination towards application and service hosting on the web which has yielded in increased demand for verification. The promising application currently available in market has witnessed to be in frequent use like mobile banking and micropayment. Unfortunately, the robustness of security as well as authentication among the parties is very slim. Moreover usability and expense of such token factor authentication device is very frustrating. One of the flaw found in mobile internet security is the complicated feeding of the password which is due to the restrictions of their input interfaces.

The aim of this paper is not only concentrated to find out certain solutions, but also concentrated on designing specification practicable information security techniques or products for the application and mobile service of cloud computing. In this paper we propose using observed user behavior to authenticate users, an approach we refer to as implicit authentication. Exclusive verification can be used to meet the following general verification requirement: 1) Used as a dual factor for client verification, exclusive authentication can supplement passwords to achieve higher-assurance verification in a cost-effective and user-friendly manner. 2) Used as a major method of verification, inherent verification can substitute passwords altogether, relieving users from the weight of feeding passwords. 3) A tertiary use of the expertise is to provide supplementary guarantee for credit card transactions, based on the security posture of the device owner.

In Section 2, we will discuss about the previous research work in this area followed by Section 3 about mobile commerce system over wireless network. Section 4 highlights about the key issues followed by security requirement in section 5. In depth discussion of micropayment system is discussed in section-6 followed by proposed system in section-7. Section 8 highlights the gaining algorithm followed by performance analyzation in section-9 and conclusion in section-10

## II. RELATED WORK

Rafael [9] presents a performance analysis of different mobile payment protocols. The performance analysis includes the computational cost required by each entity to perform all the cryptographic operations and the transmission time required to transmit each message. But the work did not considered analysis using mobile payment protocols using elliptical curve cryptography. Suresh Chari e.t. al. [10] have identified some frameworks and their inherent exposures in security issues in m-commerce. Alia Fourati [11] has worked on secure and fair auction over adhoc network. Even in this work also, some specific security issues to adhoc networks were not treated. Osman [12] presents a fully distributed and self-organizing approach to managing group membership in such a loose trading community. It is designed to suit the dynamic nature of ad hoc wireless networking and the social characteristics of ad hoc m-commerce.

Zhi-Yuan Hu [13] has designed an innovative and practical authentication system, Anonymous Micropayments Authentication (AMA), is designed for micropayments in mobile data network. But his work has a relative drawback for common problems of authentication mechanism based on symmetric key cryptography.

Xiaoling Dai [14] has researched on micropayment protocols in offline with multiple vendors.

Min-Shiang [15] has introduce several micro-payment schemes based on one-way hash chain and review some literatures on supporting multiple payment. The author has also proposed a new micropayment scheme, which achieve the following three goals: micro-payment multiple transactions, service providers, and anonymity.

Samad [16] has proposed a trust model from user point of view and combined it with MR2 micropayment scheme and called the new scheme TMR2. This trust model is supported by micropayment provider and assures the users that they will not be charged for in case the product is not satisfactory or it is corrupt.

Sung-Ming e.t. al [17] has studied various probabilistic micropayment Scheme shows that the scheme by Rivest may reduce the administrative cost of the bank, however it brings extensive computational overhead to the merchant.

Lih-Chyau Wuu [18] has proposed a secure and efficient off-line micro payment scheme which uses coin chain technique to make coin that the verification of coin can be done quickly by hash computation. This scheme also ensures that the coins could only be used by their owner, and protects the privacy of the consumer.

Vivek Katiyar e.t. al. [19] has discussed about role of Elliptical Curve Cryptography and presents a survey on the current use of ECC in the pervasive computing environment.

Husna Osman and Hamish Taylor [20] has discussed three key design considerations in implementing a fully distributed reputation system for ad hoc m-commerce trading systems, namely relevant reputation information, its storage and reliability.

Fouzia Mousumi and Subrun Jamil [21] has described cost effective push pull services officering SMS based mobile banking concept has been illustrated for 24 hours banking convenience which helps customers stay on top of any recent changes made in their current or deposit account or loan through SMS.

Arogundade e.t. al. [22] propose an open network system which can adapt to users changing needs as well as allowing effective and secured transaction via any customers' bank account.

Partha e.t. al [23] proposes a novel approach by utilizing cancelable biometric features for securely storing the fingerprint template by generating Secured Feature Matrix and keys for cryptographic techniques applied for data Encryption or Decryption.

Mohammad Al-Fayoumi [24] discuss an important e-payment protocol namely pay-word scheme and examine its advantages and limitations, which encourages the authors to improve the scheme that keeps all characteristics intact without compromise of the security robustness

Kaylash Chaudhary e.t. al [25] have carried out an assessment of micro-payment against a non-micro-payment credit systems for file sharing applications.
Charles K. Ayo and Wilfred Isioma Ukpere [26] propose a unified (single) smart card-based ATM card with biometric-based cash dispenser for all banking transactions

Wang [27] proposes a novel payment system with smart mobile devices, wherein customers are not limited to purchase e-cash with the fixed face-value
Obviously it can be seen that majority of the work is carried on wired network with much less consideration of wireless network. The issues related to dynamic topologies of wireless adhoc network is not discussed in detailed in any of the researches described above.

## III. MOBILE COMMEREC SYSTEM

The adhoc mobile commerce takes place between multiple numbers of nodes which are in proximity of each other without relying on the services of any infrastructure [28] which is very different from infrastructure mobile commerce application. Such nodes can be termed as peer node which can cooperate and participate in communication process by using their normal local resources along with their neighbor's independent on any support provided by a network service provider in order to achieve the transaction or such related task. So, following are the inherent properties of mobile commerce in adhoc network:

i. Independent of Service Provider: As the adhoc wireless network will not have a network service infrastructure and are self-organized, a dedicated service provider cannot be entrusted for allocation of maintenance task for enabling security parameters or payment scheme reliability for m-commerce applications.

ii. Restricted Scope of Communication: With various restriction in communication range especially in IEEE 802.11 [29] [30] [31] poses a challenges in adhoc networks where the network topology is normally dynamic rendering less trust on any third party service by communicating peer node to support security and/or payment in real-time application among the peers.

iii. Inadequate online time: Due to finite energy cycle and the dynamic topology of mobile devices as well as intermittent network disconnections [32], there is a restricted time during which these mobile devices can be online, which actually limits them from participating in lengthy and complex operation processes related to transactions [33]. This fact represents that complex secure operation in m-commerce need to be completed in a fairly short period and should only comprise a few simple stages if they are to have a good chance of success in terms of security.

iv. Impulsive choice in Adhoc configurations: As the adhoc wireless network has self-organizing attribute [33] which allows client that are equipped with mobile devices to instinctively participate in m-commerce transactions when the requirement arises while they are on the mobile mode.

v. Cost Effectiveness: There is no extra complicated device [34] in mobile commerce application in order to perform security operations in m-commerce over an adhoc wireless network as peer nodes which will formulate the network.

vi. Privacy: The mobile commerce application enabled in adhoc wireless network is very much appropriate for maintaining or safeguarding the privacy protocol for commercial transactions where the clients (traders) will not look for disclosing the commercial transaction information to some external entities [35]. There is no third entity which needs to be involved in order to realize the network communication.

IV. KEY PROBLEMS (3)

The serious issues to be highlighted are that conducting m-commerce secure operations over mobile adhoc wireless networks introduces added challenges and concern. Along with this, the adhoc wireless networks have particular problems which needed to be considered in research works in future. The major issued found are illustrated as below:

i. Transaction Management: It is very difficult to execute secure effective transaction methods and moreover updates in mobile adhoc networks, which is due to its sole distinctiveness e.g. lack of infrastructure, having a dynamic network topology and using resource constrained devices. Majority of the traditional research work has utilized infrastructure based m-commerce which depends on a client/server model where information is fundamentally located placed on servers within the wired network and peer nodes act as clients accessing the services provided by the servers [36] along with an issue of service unavailability due to network disconnections [36]. Also, the in-depth idea of a transaction can be difficult to enforce as network intermittent disconnections will affect a particular service in a secure m-commerce operations succession to fail and accordingly the secure connectivity would be considered unfinished and will be subjected to abort [37].

ii. Delivery of Service: Due to the unique characteristics and complexities of an adhoc wireless network, existing service discovery and delivery protocols [38] do not seem to suit the needs of an adhoc wireless network, making them unsuitable for m-commerce oriented scenarios. Service advertisements and deliveries may need to be disseminated by a mix of a store and forward strategy as well as local multicasting to cope with intermittent online connectivity [38].

iii. Trust-System: One of the important factor of online communication in terms of security will be Trust, which assists in participating entities to ensure the secure transaction by extenuating improbability and risks involved in the transactions, such as ambiguity about trading groups or entities' pattern in fulfilling the transaction agreements [39]. On the other hand, as mobile adhoc network cannot rely on a network service provider to facilitate security services such as certification authority (CA) which can assists to design trust system among peer nodes in the existing network. It can also be observed that peer nodes have to rely on their peers in the network to provide trust verification in order to evaluate other nodes' fidelity. Yet, the nature of an adhoc wireless network makes trust scheme founding in this network intricate to accomplish.

V. SECURITY REQUIREMENTS

In order to establish an appreciably tenable and trusted atmosphere for an operation to take place as well as to facilitate self-reliance to trading entities to participate in secure m-commerce operations, the following security services can be considered as an important functional requirement.

i. Authentication: Authentication is the first step which facilitates both trading entities participated in m-commerce transactions to substantiate the identity of each other before any transaction is conducted among the groups. This service ensures that any illegal third party or external agents is not masquerading as a legitimate party.

ii. Privacy: Privacy guarantees that secure transaction information sent across the network is incomprehensible by illegal third parties such as malicious eavesdroppers or peers acting only as communication relays, or DDoS attacks.

iii. Reliability: Reliability ensures that a message is being transferred is not illegitimately altered or destroyed during the transmission without this being detectable at the receiver side of an m-commerce system.

iv. Non-repudiation: This is another property of which assures that if an entity transmits a message, it will not be able to move away with denying after sending the message. Usually in m-commerce transactions, neither sender nor receiver should realistically be able to renounce offers or bargains struck between them. The sender should not be able convincingly to deny having sent the transaction message and the receiver should be able to verify that the transaction message can only have been sent by the specified sender and thus able to prove that a business has taken place between them. Along with this, as m-commerce transactions will include the threat of misbehavior among the trading entities, which they need support in measuring the intensity of reliability of other trading entities. Hence, attestation is another important security service for an adhoc m-commerce.

v. Attestation: Attestation enables an adhoc m-commerce peers to vouch for the identity, trading history or transaction reputation of other peer nodes. It assists alleviate threat in transacting with formerly unknown entities.

## VI. ABOUT MIROPAYMENT SYSTEM

A micropayment is a financial transaction involving a very small amount of money and usually one that occurs online [40]. One problem that has prevented their emergence is a need to keep costs for individual transactions low which is impractical when transacting such small sums even if the transaction fee is just a few cents [24]. Micropayments have to be appropriate for the transaction of non-tangible merchandise over the Internet which inflicts necessities on speed and cost of processing of the payments: delivery occurs nearly immediately on the Internet, and often in arbitrarily small pieces. On the other hand, the bottleneck in sales of tangible merchandise, management and distribution, sets a lower bound particularly for costs to remain economical. So, the evaluation criteria of micropayment systems should include [41]:

Ease of use: The application must be easy to use for the clients. There is no authorization login and PIN number to be fed all the time. The customer only needs to click and to buy a page in the web page with a micropayment system in a few seconds.

Security: The aim of security in the payment procedures is to prevent any group from cheating the system. For customers and external adversaries the forms of cheating security, which are detailed to payment design, are extra expenditure of coins and creation of false coins forgery during payment.

Anonymity: The customer anonymity should be protected. An elementary property of physical cash is that the association between customers and their purchases is untraceable. This means that the payment systems do not allow payments to be traced without compromising the system's security. This may encourage some potential customers to start using the payment system.

Divisibility: The protocol supports multiple denominations and a range of payment values.

Performance: The protocol provides high-volume payment support.

Robustness: The protocol is tolerant of network bottlenecks and broker/authorizer down-time.

Table 1. Comparison of E-commerce payment methods

| Property | CyberCash [41] | MPay [41] | PayWord [41] | NetPay [41] |
|---|---|---|---|---|
| Ease of Use | Low | High | Medium | High |
| Security | High | Medium | Low | Medium+ |
| Anonymity | Low | Low | Low | Medium+ |
| Divisibility | Very High | Very High | High | High |
| Performance | Very Low | High | Medium | Very High |
| Robustness | Low | High | High | High |

There is a growing need for an effective, efficient micro-payment technology for high-volume, low-value E-commerce products and services. Current macro-payment approaches do not scale to such a domain. Most existing micro-payment technologies proposed or prototyped to date suffer from problems with security, lack of anonymity and performance.

## VII. PROPOSED SYSTEM

The proposed system assumed the attacker or intruder which has an access to the confidential physical information to the trusted hand held device which the user normally use for mobile banking or performing very confidential transaction. In the current 3G enabled cell phone, normally all the phone comes with an inbuilt application of Facebook, Twitter, or some other application like mCheck of Airtel. Therefore, the intruder might also be interested in other alternative resource which is connected with the trusted hand held device like accessing the account details from phone book or message archives etc. Not only this, the presence of an type of malicious applications could also pose a huge threat towards the device and the premium services associated with the device. Example of such attack could be cloning attack where the malicious program will attempt to send information on the victim trusted handheld device to the colluding user available in the network and then slowly poison the network sending certain information of services which by default is supposed to create an event after it. Such types of lethal threat can be prevented by considering the message packets being digitally signed by the subscriber identity module card which will be most difficult to be duplicated. The consistent malicious programs on the mobile phones will able to have more lethal effect as the complete control structure will be robustly created by the attacker for which every event performed from the device can be monitored by the intruder. It also raises possibility of duplication of the action performed by the genuine user by the fraudulent user. Not only this various availability of spy softwares will increase the threat exponentially.

The proposed solution against such types of issues is highlighted in this paper as a system knowledge acquiring algorithm. The proposed system initially acquires knowledge from a client's system from their previous actions

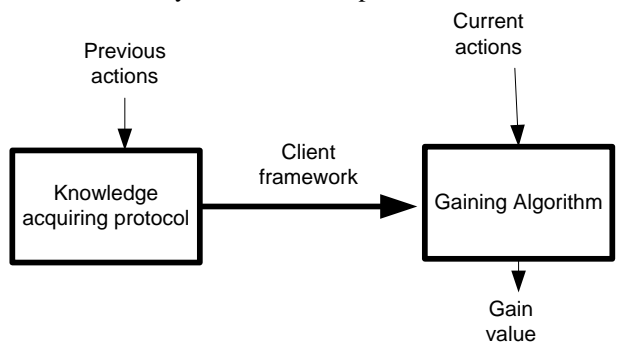

Fig 1.Proposed Architecture

. In order to design a robust verification assessment in real-time scenario, the proposed system uses a gaining protocol which assess the client system and their recent actions performed in previous history of transactions or any other activities over the device and then it yields a gain value identifying the probability that the genuine client is utilizing the trusted hand held device. The gain value is considered to design a verification decision which characteristically uses a threshold factor in order to choose whether to accept or to decline the genuine user. Not only this, the threshold factor can also deflect from diverse application, which is dependent on if the system is responsive to optimal security measure. The gain value could also be considered for a dual aspect gauge to supplement conventional secret word based authentication system, which we use currently.

The actions of the genuine client are characterized by the client system framework. An immature framework can be discussed where it can be considered about the liberty among the various diversified sections of user actions. Representing alternatively, it can also be considered that the client's trusted hand held device is free from their location, usage of the service, as well as any other activities. The framework assumed the client's actions performed is completely dependent on the instant of the time in day as well as week, probably can be month too. To cite an example, it can be said that one client might use both incoming as well as outgoing calls very frequent in morning but might not work out in outgoing calls in afternoon. He might get only frequent calls in afternoon. Let $F_1, F_2, ...., F_k$ represents independent arbitrary feature variable. Let is assume $F_1$ is time elapsed since the previous good calls, $F_2$ is inter arrival time between bad calls, $F_3$ represents location coordinates etc. The good call represents the incoming or outgoing calls done from the genuine user phone book and bad calls represents any incoming or outgoing calls which are not listed in the memory of the phone or a SIM card. The client's framework is a multiplication of R probability function trained on the variable T as instance of time.

Therefore client's framework is,

$$[P(F_1/T), P(F_2/T), ... P(F_R/T]$$

The knowledge acquiring protocol fundamentally computes such functions structuring client's model.

### VIII. GAINING PROTOCOL

With a facilitated client's framework and previously known set of actions of the client, the gaining protocol yields a gain value representing the probability that the trusted handheld device is under the control of genuine user. This can also be described as gaining independent chareteristics. The gaining function is developed in a very secure and robust way under the independent characteristic framework. The client current actions can be represented as tuple $(T, F_1, F_2, .., F_R)$ where current time is shown as T and F1, F2, ….FR represents the values of variables $(F_1, ... , F_R)$ at instant of time t. The significant perception for this logic is to evaluate a discreet gain for each feature and then utilize it a function in order to gather such distinctive gain values into a final evaluation. Fundamentally, we will have R gaining functions represented as $G_1, G_2, \ldots, G_R$. Let $1<i<R$. With a facilitated probability function $P(F_i/T)$ and an monitored value $f_i$, the $i^{th}$-gaining function $G_i$ yields a gain value of $g_i$ exclusively for this characteristic. Considering a gaining function for F1 as time elapsed since previous good call, the perceptual logic is that the gain should decompose over coarse of time during periods of idle situation of the trusted hand held device. However, the rate of decomposing will naturally dependent on the instant of time of day. The gain value will decrease frequently in the afternoon when there are no actions of incoming calls but lots of calls are made. Also if the client makes absolutely no calls at midnight then the gain value will decrease rather slowly over coarse of time. The gaining function for a nominal client can be represented as follows. Let f1 represent the drift since the previous call at time t. Let cumulative distribution be represented as $H(x|T=t) = Pr(F_1<x|T=t)$ of variable F1 for time t. The gain is defined to be the feasibility that a drift of f1 or longer is monitored at time t.

$$G1(f1)=1-H(f1|T=t) \qquad (1)$$

The gain function for the client position might be allocated to a position visited at a specific time of day a gain which is inversely proportional to the Euclidean distance to the nearest position group which is connected with that specific instant of the day. A client who specifically is at "office" group during office hours and at a "resident" group at night can receive the maximum gain for being positioned at an expected group at the expected time. Position near expected group would receive incomplete credit which reduces to "0" as the distance to the group increments.

The next phase will be knowledge acquiring with gaining protocol. With the facilitated distinctive gain values for the R different charecteristics, the system calls $f(G1(f1), \ldots GR(vR))$ to evaluate the final value of gain. Citing an example, let us say that each gain $Gi(f1)$ $(1<i<R)$ is the feasibility of the fi, which means $Gi(fi)=Pr[Vi=vi]$ as in equation (1). Exactly after this step, a common process to combine the gain values is to estimate the combined feasibility of $(f1, \ldots, fR)$. The final gain value will be the multiplication of these feasibilities: $f(g1, \ldots, gR)=g1.g2.. . gR$. Or else alternative feasible structure for function is a weighted addition: $f(g1, \ldots, gR):= w1.g1+w2.g2+\ldots+wR.gR$. The weights $w1, \ldots, wR$ has to be estimated through a potential training procedure. In order to acquire knowledge using gaining protocol, we need to consider that the system gather set of actions from the individual clients. This collection of information will be categorized into evaluation set and training set. The training set will comprise of utility as positive samples in the training procedure. The system in hybrid method generates attack information for training process. In specific, the system will deploy wedging procedure for creating negative samples. That will mean if a client P and client Q come into view in the proximity of each other at instant of time t, than the system wedge the information for P before t and the information for Q after time t. The proposed framework represents a hybrid process of initiating an attack model where Q picks up or intrude P's trusted hand held device and initiates using it maliciously. In practicality, Q could be any associated relation or may be completely outsider (stranger). Training the weights $w1, \ldots, wR$ can be represented as a issue of minimization which will mean that if the system fixes the rate of false negative e.g. it can be said that a genuine legal client is declined the permission of access and has to feed the password frequently in day. Therefore the proposed system aims at minimize the rate of false positive which is failure to identify an attack and the time till the detection in the presence of an attack.

## IX. PERFORMANCE ANALYZATION

The analyzation of the proposed system is done considering the mobile platforms like iPhone, Andriod as well as Symbian mobiles. The experiment is done using Andriod enabled hand held device. The proposed experiment is executed after gathering a set of actions and event for Andriod enabled device.

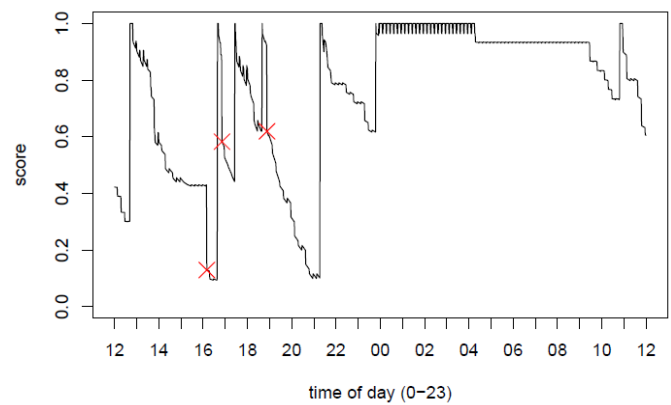

*Figure 4: Gain* over time. Each red "X" marks a call to an unknown number, where the score is decreased. [The x-axis symbolize the time of day expressed in hours 0-23. The y-axis symbolize the score, which is a value between 0 and 1]

Various sets of information like calls, emails, SMS, GPRS enabled packets stored, phone books etc are considered. Various activity of the emails are also gathered like incoming emails for read and unread message, deleted emails, classification of emails in respective folders, sent emails etc. For calls events, various incoming and out going calls are registered considering both good and bad calls. The same procedure is also applicable for SMS. Other host utilities like memory card, calendar, image, etc are also tracked. The mentioned gather information is collected in last 2 months of the phone information for each participant in the observation. The interesting observation is that their call logs represents discreet patterns depending on the instant of the time.
]

For the complete evaluation, the framework will assist the gain which will group with all the features collected. In the initial stage of analysis, a gain protocol for phone data with two features are designed. The features are F1 and F2 which represents time drifted since previous good call and quantity of successive bad calls. If the majority of the current call is good call, than F2 is 0. If the majority of the 3 calls are good, bad and bad respectively, than F2 is "2". Figure 4 will highlight a graph for verification gain for one client in last one day. It is seen that whenever the client initiates a call to a phone number listed in the phone book or the SIM card (good calls), the gain value is increased up to 1. From the figure it can be understood that the gain value decreases when the bad call is initiated. In the idle duration of a day, the gain decompose over the coarse of time. Predominantly, the graph also highlights that the gain decompose frequently in the afternoon in comparison to night. During an idle duration between 6 pm to 10 pm, the gain decomposes frequently. Also it can be noted that in an idle period between midnight to dawn, the gain value decomposes very frequently. The reason behind this is found this client specifically initiates maximum calls between 6 pm and 10 pm, but will not initiate any calls in midnight. The minimum limited frequency is due to an object in the gain protocol defined in equation 1. But this does not assure that the gain value is strictly reducing. It might be feasible to have minimum confined frequency while the gain will decompose over a longer duration of instant of time.

The experiment is also conducted by analyzing the gain protocol on an attacker trace acquired by wedging two client's trace. It was seen that the gain value of the attacker frequently reduces to 0 as the attacker labels the disjoint set of numbers from the holder of the trusted handheld device.

X. CONCLUSION

In the proposed system, a framework for hidden authentication scheme is highlight, which assures better and robust security for mobile commerce application in existing scenario of usage in cloud. Majority of the existing research are focused on designing application for security in m-commerce application, but ignorance in identifying the potential threat caused due to normal usage of mobile phone with premium services is highlighted in this paper. The designed application is tested by gathering real time data from android enabled hand held device and tested using our proposed gaining protocol to identify the risk of attack from the illegal clients who malicious wants to access the resources of the genuine client for harmful intentions.